\newcommand{\be}{\begin{equation}}
\newcommand{\ee}{\end{equation}}
\newcommand{\bear}{\begin{eqnarray}}
\newcommand{\eear}{\end{eqnarray}}
\newcommand{\ba}{\begin{array}}
\newcommand{\ea}{\end{array}}

\newcommand{\vev}{v}

\documentclass[twocolumn,showpacs,nofootinbib,preprintnumbers,amsmath,amssymb,superscriptaddress]{revtex4}
\usepackage{epsfig,graphicx,psfrag}

\renewcommand{\theequation}{\thesection.\arabic{equation}}

\begin{document}

\preprint{\small FERMILAB-PUB-06-439-T, \ \ NUHEP-TH/06-07, \ \ UCI-TR-2006-19} \ \ \ 
\preprint{\small hep-ph/0612017\rule{0em}{1.3em}}  

\title{\Large \bf Gauge Trimming of Neutrino Masses }

\author{Mu-Chun Chen}
\affiliation{Theoretical Physics Department,  Fermilab, Batavia, IL 60510, USA} 
\affiliation{Department of Physics \& Astronomy, University of California, Irvine, CA 92697, USA}

\author{Andr\'e de Gouv\^ea}
\affiliation{Northwestern University, Department of Physics \& Astronomy, 
Evanston, IL~60208, USA}
\affiliation{Theoretical Physics Department,  Fermilab, Batavia, IL 60510, USA}

\author{Bogdan A. Dobrescu}
\affiliation{Theoretical Physics Department,  Fermilab, Batavia, IL 60510, USA}

\date{December 1, 2006 
\rule{0em}{1.8em}} 

\begin{abstract}\rule{0em}{1.8em}
We show that under a new $U(1)$ gauge symmetry, which is non-anomalous in the presence of one ``right-handed neutrino'' per generation and consistent with the standard model Yukawa couplings, the most general fermion charges are determined in terms of four rational parameters. 
This generalization of the $B-L$ symmetry with generation-dependent lepton 
charges leads to neutrino masses induced by operators of high dimensionality. 
Neutrino masses are thus naturally small 
without invoking physics at energies above the TeV scale, 
whether neutrinos are Majorana or Dirac fermions.
This ``Leptocratic'' Model predicts the existence of 
light quasi-sterile neutrinos with consequences for cosmology, and 
implies that collider experiments may reveal the origin of neutrino masses.

\end{abstract}

\pacs{12.60.Cn,14.60.Pq}

\maketitle

\setcounter{equation}{0}\setcounter{footnote}{0}
\section{Introduction}
\label{sec:intro}

It is known, beyond any reasonable doubt, that at least two of the 
three known neutrinos have nonzero masses \cite{TASI, strumia_vissani}. 
While nonzero, the neutrino masses are at least one million 
times smaller than the electron mass. 
With the standard model field content, neutrino masses arise only 
from nonrenormalizable operators, of the type $\overline{\ell}_L^c \ell_L H H$,
where $\ell_L$ is a lepton doublet and $H$ is the Higgs doublet \cite{Weinberg:1979sa}.
These dimension-5 operators are suppressed by a mass scale $\Lambda$ that must be 
of order $10^{14}$ GeV if the dimensionless 
coefficients of the operators are of order unity.  

Another way of rendering the neutrinos massive is to postulate the existence of 
gauge-singlet chiral fermions, $n_R^k$, where $k$ is a flavor index.
These interact with lepton doublets through Yukawa interactions 
involving the Higgs doublet, so that electroweak symmetry breaking leads to 
Dirac masses for the neutrinos.
The $n_R^k$ fermions, usually referred to as `right-handed neutrinos', have
Majorana mass terms allowed by the electroweak symmetry.
If the Majorana masses are larger than the Dirac ones, 
then  integrating out the right-handed neutrinos generates the 
dimension-5 operators mentioned above, with the scale $\Lambda$ given by a Majorana mass and
the dimensionless coefficients given by products of Yukawa couplings. 
This renormalizable model yields small neutrino masses even for large Yukawa 
couplings \cite{seesaw, theory_reviews}, provided the Majorana masses are 
orders of magnitude above the electroweak scale.

In this paper we explore an alternative explanation for the smallness of 
neutrino masses, which does not require physics at energy scales much above 1 TeV.
We extend the standard model by including right-handed neutrinos and a 
non-anomalous $U(1)$ gauge symmetry, referred to as $U(1)_{\nu}$.
The lepton and Higgs charges can be chosen such that all neutrino Yukawa 
couplings and right-handed Majorana mass terms are not  gauge 
invariant under $U(1)_{\nu}$. Thus, neutrino masses (both active and sterile) may be
generated only when the new gauge symmetry is spontaneously broken. 
It turns out that in many cases the neutrino masses arise only
through operators of  mass dimension six or higher, so that 
they are naturally small even when 
the ultraviolet cutoff of the $U(1)_{\nu}$-extended standard model is
as low as the TeV scale.

The anomaly cancellation conditions strongly restrict the choices of 
$U(1)_{\nu}$ charges for leptons.
As a result, this mechanism is predictive, and it leads 
to distinct phenomenological consequences.
If the lepton charges are generation-dependent, then 
the $U(1)_{\nu}$ symmetry serves as a natural flavor symmetry, and may
help explain the pattern of lepton mixing.
The right-handed neutrinos are likely to behave as sterile neutrinos.
These light, weakly coupled, electroweak-singlet 
fermions are similar to the ones present in the case of a 
very low-energy Type-I seesaw \cite{deGouvea:2005er}, playing a role 
in neutrino oscillations,
searches for neutrino masses, or astrophysics \cite{deGouvea:2006gz}. Light
sterile neutrinos have also been identified as a dark
matter candidate \cite{Asaka:2005an}.
The interesting feature in our case is that the sterile neutrinos are 
only partially ``sterile'', because they interact with ordinary matter via exchange of the 
$Z^\prime$ gauge boson associated with the $U(1)_{\nu}$ symmetry. 
In particular, even when the Majorana masses vanish and 
the right-handed neutrinos pair up with the left-handed ones
to form Dirac neutrinos, the new light fermionic degrees of freedom have a variety 
of observable effects.

Our models lead to interesting correlations between neutrino physics and collider searches at 
the TeV scale. For example, if a $Z^\prime$ boson is discovered and its couplings are determined at the LHC
and ILC, then the structure of the operators responsible for neutrino masses may be 
revealed.
Moreover, if the ultraviolet cutoff for the $U(1)_{\nu}$-extended standard model is close
to the TeV scale, other new physics effects are likely to be within the reach of 
upcoming collider experiments.

The solutions for $U(1)_{\nu}$ charges presented here are generalizations with 
generation-dependent lepton charges of the nonexotic $U(1)$ gauge symmetry discussed
in Ref.~\cite{Appelquist:2002mw,Ferroglia:2006mj}. 
In the particular case where only the right-handed neutrinos carry $U(1)$
charges, the symmetry breaking scale may be very low, leading also to light sterile neutrinos 
\cite{Davoudiasl:2005ks, Sayre:2005yh}.
If new exotic fermions charged under the standard model gauge group are included, then 
there are more general $U(1)$ charge assignment with implications for 
neutrinos (see, {\it e.g.}, \cite{Kang:2004ix, Barr:1986hj,Carena:2004xs}). 
It is straightforward to extend our models to include supersymmetry
(for example, the case without charged right-handed neutrinos is analyzed in \cite{Demir:2005ti}).
In contrast to our non-anomalous $U(1)_\nu$ broken at the TeV scale, in models where
the $U(1)$ anomalies are cancelled by the Green-Schwarz mechanism
the constraints on fermion charges are greatly
simplified at the cost of pushing the $U(1)$ breaking scale close to the 
fundamental string scale ~\cite{Ibanez:1994ig,Dreiner:2006xw}.


In Sec.~\ref{sec:formalism} we present general features of the $U(1)_{\nu}$ extension on the standard model, 
concentrating on its consequences for neutrino masses. In Sec.~\ref{sec:model} we solve the anomaly cancellation 
conditions in the case of one ``right-handed neutrino'' per generation,
and describe some  concrete examples that lead to interesting neutrino phenomenology.
Consequences of relaxing various assumptions are discussed in
Sec.~\ref{sec:cases}. Cosmological constraints  
and possible collider probes of these models are discussed in Secs.~\ref{sec:cosmo} and \ref{sec:collider}, respectively. 
Our concluding remarks are presented in Sec.~\ref{sec:end}. 
A few technical issues regarding the approximate diagonalization of a $6\times 6$ neutrino 
mass matrix have been relegated to an Appendix.

\setcounter{equation}{0}
\setcounter{footnote}{0}
\section{$U(1)$ Extension of the Standard Model Gauge Group}
\label{sec:formalism}

We extend the standard model gauge group by a new $U(1)$ symmetry, labeled $U(1)_{\nu}$.
We assume that the {\it quark} charges under the $U(1)_{\nu}$ gauge symmetry 
are generation independent, so that we do not have to
worry about large contributions to flavor-changing neutral currents from 
tree-level exchange of the $U(1)_{\nu}$ gauge boson~\cite{Langacker:2000ju}.
The left-handed $SU(2)_L$-doublet quarks, $q_L^i$, where $i=1,2,3$ labels the generations, 
and right-handed $SU(2)_L$-singlet quarks, $u_R^i, d_R^i$,
have $U(1)_{\nu}$ charges $z_q, z_u$ and $z_d$, respectively.

In the case of the leptons, we allow generation-dependent $U(1)_{\nu}$ charges. We discuss the implications of 
the non-observation 
of flavor-changing neutral currents involving electrons, muons or taus in Sec.~\ref{sec:formalism-fcnc}.
The left-handed $SU(2)_L$-doublet leptons, $\ell_L^i$, and right-handed $SU(2)_L$-singlet leptons, $e^i_R$,
have $U(1)_{\nu}$ charges $z_{\ell_i}$ and $z_{e_i}$, respectively.
We also extend the standard model particle content to include some fermions $n_R^k$, $k=1,\ldots, N$ that have
charges $z_{n_k}$ under $U(1)_{\nu}$ and  are singlets under the standard model gauge group.  
Note that the total number of these  right-handed neutrinos,
$N$, may be any positive integer, as it is not necessarily related to the number of quark and lepton generations.

Besides the standard model Higgs doublet $H$, of hypercharge +1 and charge 
$z_H$ under $U(1)_{\nu}$, we introduce a scalar field, $\phi$,
which is charged under $U(1)_{\nu}$ and is a singlet
under $SU(3)_c\times SU(2)_L\times U(1)_Y$. We 
assume for simplicity that the vacuum expectation value (VEV) of $\phi$
is the only source of $U(1)_{\nu}$ breaking when $z_{H} = 0$. 
A larger $U(1)_{\nu}$ breaking sector may lead 
to different phenomenological consequences, as briefly discussed
in Sec.~\ref{sec:cases}.

\subsection{Anomaly cancellation and masses for electrically-charged fermions}
\label{sec:chargedmasses}

The $U(1)_{\nu}$ fermion charges are constrained by the requirement of anomaly 
cancellation. 
The $[SU(3)_c]^2U(1)_{\nu}$ and $[SU(2)_L]^2U(1)_{\nu}$ anomalies cancel only if 
\begin{eqnarray}
z_d &=& 2z_q-z_u ~, \nonumber \\ [0.3em]
z_{\ell_3} &=& -(9z_q+z_{\ell_2}+z_{\ell_1}) ~. 
\label{eq:zl3}
\end{eqnarray}
Using these equations, the $[U(1)_Y]^2U(1)_{\nu}$ and $U(1)_Y [U(1)_{\nu}]^2$ 
anomaly cancellation conditions take the following form, respectively:
\begin{eqnarray}
\sum_{i=1}^3 z_{e_i} &=& -3\left( 2z_q + z_u \right) ~, \label{eq:y2nu1} \\ [0.5em]
\sum_{i=1}^3 \left(z_{\ell_i}^2 - z_{e_i}^2\right) 
&=& 3\left( z_q - z_u \right)\left( 5z_q + z_u \right) ~. 
\label{eq:y1nu2}
\end{eqnarray}
Finally, the $U(1)_{\nu}$ gauge-gravitational anomaly and the
$[U(1)_{\nu}]^3$ anomaly cancel only if 
\begin{eqnarray}
\sum_{k=1}^N z_{n_k} & = & -3\left( 4z_q - z_u \right) ~, \label{linear} \\ [0.4em]
\sum_{k=1}^N z_{n_k}^3 & = & \sum_{i=1}^3 \left( 2 z_{\ell_i}^3 - z_{e_i}^3\right) 
- 54 z_q \left( z_q - z_u \right)^2 \label{cubic} ~. 
\end{eqnarray}

Another set of constraints on the $U(1)_{\nu}$ charges comes from
the requirement that there are terms in the Lagrangian which generate
the observed quark and lepton masses.
The large top-quark mass implies that its Yukawa coupling to the Higgs doublet is 
gauge invariant, so that 
\begin{equation}
z_H=z_u-z_q ~.
\label{eq:zh}
\end{equation}
This, combined with the anomaly constraints and the quark charge universality discussed above, 
implies that all standard model Yukawa couplings involving quarks are allowed by all gauge symmetries.

We further require that the ``diagonal'' Yukawa interactions involving the tau and the muon,
\begin{equation}
\lambda_e^{ii} \overline{\ell}_L^i e_R^i H ~,~~i= 2,3,
\label{cc_mass}
\end{equation}
are gauge invariant. The entries of the $3\times 3$ matrix $\lambda_e$ are
dimensionless Yukawa couplings.
Using Eqs.~(\ref{eq:zh}) and (\ref{eq:zl3}), we find that these two conditions are satisfied if
\begin{eqnarray}
z_{e_2} &=& z_{\ell_2}-z_u+z_q ~, 
\nonumber \\ [0.3em]
z_{e_3} &=& -\left( 8 z_q + z_u +z_{\ell_1} +z_{\ell_2} \right) ~.
\label{eq:e2e3}
\end{eqnarray}
Consequently, the $[U(1)_Y]^2U(1)_{\nu}$ anomaly cancellation condition 
given in Eq.~(\ref{eq:y2nu1}) implies 
\begin{equation}
z_{e_1}= z_{\ell_1}-z_u+z_q ~, 
\label{eq:e1}
\end{equation}
so that the diagonal electron Yukawa interaction is automatically gauge invariant.
It is interesting that the conditions $z_{\ell_{i}}-z_{e_{i}} = z_{q}-z_{u}$ for $i=1,2,3$
reduce the quadratic equation Eq.~(\ref{eq:y1nu2}) to a linear one, 
which is automatically satisfied once the $[SU(2)_{L}]^{2} U(1)_{\nu}$ 
and $[U(1)_{Y}]^{2} U(1)_{\nu}$ anomaly cancellation conditions 
given in Eq.~(\ref{eq:zl3}) and (\ref{eq:y2nu1}) are satisfied. 
Note that off-diagonal, renormalizable charged-lepton Yukawa couplings may 
be forbidden by $U(1)_{\nu}$ invariance. Similar to the quark sector, the hierarchy 
of masses for the electrically-charged leptons  is generically dictated by the entries of 
$\lambda_e$, which do not depend on the existence of $U(1)_{\nu}$. 
On the other hand, lepton mixing can be quite sensitive to the choices 
of the $z_{\ell_i}$ and $z_{e_i}$ charges, as discussed in more detail below.

Altogether, the nine $U(1)_{\nu}$ charges of the standard model fermions 
may be expressed in terms of only four of them, chosen in Eqs.~(\ref{eq:zl3}),  
(\ref{eq:e2e3}) and (\ref{eq:e1}) to be
$z_q$, $z_u$, $z_{\ell_1}$ and $z_{\ell_2}$. Furthermore, the  $U(1)_{\nu}$ charges of the 
$N$ right-handed neutrinos are not independent: one of them may be eliminated using 
the $U(1)_{\nu}$ anomaly cancellation condition [see Eq.~(\ref{linear})], and the 
remaining $N-1$ charges must satisfy the cubic equation (\ref{cubic}). 

All $U(1)$ charges must be commensurate, {\it i.e.}, their ratios are 
rational numbers\footnote{If the charges are not commensurate, then 
the $U(1)$ group does not have the topology of a circle, but rather that 
of an infinite line.}. 
This is necessary so that low-energy theories containing $U(1)$ gauge
groups can be embedded in theories which are well-behaved in the ultraviolet.
By changing the normalization of the gauge coupling, the commensurate charges may 
always be taken to be rational numbers or even integers. For the rest of the paper 
we will treat all $U(1)_{\nu}$ charges as rational numbers. 
It follows that the  cubic equation (\ref{cubic}) has solutions only for 
very special choices of the $z_q$, $z_u$, $z_{\ell_1}$, $z_{\ell_2}$  and $z_{n_i}$ 
charges. 

It has been shown in Ref.~\cite{Batra:2005rh} that by including a sufficiently large 
number of right-handed neutrinos, the anomaly conditions may always be solved.
However, if too many right-handed neutrinos were included, then $U(1)_{\nu}$ 
would become strongly coupled at a low energy scale, 
and the model would be phenomenologically ruled out.
In this paper we tackle the inverse problem: fixing the number of  
right-handed neutrinos, we seek rational solutions to the anomaly cancellation 
conditions, and then study the phenomenological consequences
of the few cases where we find solutions.

\subsection{Neutrino mass terms}

For random choices of the  $U(1)_{\nu}$ charges, the usual 
neutrino mass terms, of the type $\overline{\ell^c}_{L}^{i} \ell_{L}^{j}HH$ or 
$\overline{\ell}_{L}^{i} n_{R}^{k}\tilde{H}$ ($\tilde{H}\equiv i\sigma_2H^{*}$), 
are likely to be forbidden by $U(1)_{\nu}$ invariance. However, 
higher-dimensional operators involving the scalar $\phi$ may generate neutrino 
masses while being $U(1)_{\nu}$ invariant.
The leading operators that yield neutrino masses are
\begin{eqnarray}
&& \hspace*{-1.4em} \sum_{i,j} 
\frac{c_\ell^{ij}}{\Lambda} \left(\frac{g_\phi}{\Lambda} \phi\right)^{q_{ij}}
\!\overline{\ell^c}_{L}^{i} \ell_{L}^{j}HH
+\sum_{i,k}\lambda_{\nu}^{ik} \left(\frac{g_\phi}{\Lambda} \phi\right)^{p_{ik}}
\!\overline{\ell}_{L}^{i} n_{R}^{k}\tilde{H}
\nonumber \\
&& \hspace*{-1em} 
+ \sum_{k,k'} c_n^{kk'} \Lambda \left(\frac{g_\phi}{\Lambda} \phi\right)^{r_{kk'}} 
\!\overline{n^c}_{R}^{k} n_{R}^{k'} + {\rm H.c.} ~,
\label{general_mass}
\end{eqnarray}
where  $\lambda_{\nu}$, $c_\ell$, $c_n$ are dimensionless couplings, 
$i,j=1,2,3$, $k,k'=1,\ldots , N$, and 
$\Lambda>\langle\phi\rangle$ is the energy scale above which the theory is no longer valid.
The dimensionless parameter $g_\phi$ accounts for the fact that each $\phi$ field participating 
in a higher-dimensional operator is presumably associated in an underlying renormalizable 
theory with a heavy fermion exchange, of typical mass $\Lambda$ and typical Yukawa
coupling $g_\phi$. 
The expressions above are allowed only for integer values of 
$p_{ik}$, $q_{ij}$, and $r_{kk'}$. If any of these integers is negative, 
then $\phi$ needs to be replaced by $\phi^\dagger$ in the corresponding operator, and 
the absolute value of the exponent is to be used. 

The exponents $p$, $q$ and $r$ can be expressed as a function of 
the $U(1)_{\nu}$ fermion charges as
\begin{eqnarray}
p_{ik} &=& z_u-z_q+z_{\ell_i}-z_{n_k} ~,\nonumber \\ [0.3em]
q_{ij} &=& 2(z_q-z_u)-z_{\ell_i}-z_{\ell_j} ~,\nonumber \\ [0.3em]
r_{kk'} &=& -z_{n_k}-z_{n_{k'}}~,
\label{eq:exp-gen}
\end{eqnarray} 
where we have normalized the gauge coupling such that the $\phi$ charge is +1.
The charges that determine the exponents above are not all independent, 
and there are correlations among the  different exponents. 

Replacing the $\phi$ and $H$ fields by their VEVs, we 
find the following mass terms for the three active and $N$ sterile neutrinos:
\begin{eqnarray}
&& \vev \sum_{i,k} \lambda_{\nu}^{ik} \epsilon^{|p_{ik}|} 
\overline{\ell}_{L}^{i} n_{R}^{k} 
+ \frac{\vev^2}{\Lambda} \sum_{i,j} c_\ell^{ij} \epsilon^{|q_{ij}|} \overline{\nu^c}_{L}^{i} \nu_{L}^{j} 
\nonumber \\ [0.3em]
&& \hspace*{3em} 
+ \Lambda \sum_{k,k'} c_n^{kk'} \epsilon^{|r_{kk'}|} \overline{n^c}_{R}^{k}  n_{R}^{k'} + {\rm H.c.} ~,
\label{nu_mass}
\end{eqnarray} 
where $\vev \simeq 174$ GeV is the magnitude of the Higgs VEV, and 
\begin{equation}
\epsilon\equiv g_\phi \frac{\langle\phi\rangle}{\Lambda} ~,
\label{epsilon}
\end{equation}
is a small parameter ($\epsilon \ll 1$) that plays an important role in what follows.
We assume for simplicity that $\langle\phi\rangle>\vev$. Lower values for 
$\langle\phi\rangle$ are possible, but the limits are highly model dependent. 

Schematically, the $(3+N)\times(3+N)$ neutrino mass matrix $M_{\nu}$ is given by
\begin{equation}
M_{\nu}\sim\left(\begin{array}{c|c} \displaystyle
\frac{\normalsize \vev^2}{\normalsize \Lambda} \epsilon^{|q|} 
& \vev \epsilon^{|p|} \\ & \\ \hline & \\ \vev (\epsilon^{|p|})^{\top} 
& \Lambda \epsilon^{|r|} \end{array}\right) ~~.
\end{equation}
Mostly-active neutrino masses arise through a combination of the 
$\overline{\nu^c}_{L} \nu_{L}$ mass terms
(sometimes referred to as the Type-II 
seesaw contribution) and the (Type-I) seesaw contribution, proportional to 
$\epsilon^{2|p| - |r|}$ if the Dirac masses are much 
smaller than the right-handed Majorana masses.

It is curious that a pure Type-I seesaw ({\it i.e.}, vanishing $\epsilon^{|q|}$ 
term) is not 
attainable, provided that all gauge invariant operators are present. 
To see this, note that if there is a Type-I seesaw contribution 
to the active neutrino mass, then there are integer values for $p_{ik}$ and 
$r_{kk'}$, implying that 
the same is true for\footnote{The converse is not necessarily true: it is
possible to choose charges so that $q_{ij}$ is an integer for some $i,j$ pair, 
while there are no integer
$p_{ik}, r_{kk'}$ for all values of $k,k'$ and $i$. In this case, only the 
Type-II contribution is present. } 
\begin{equation}
q_{ij} = -p_{ik}-p_{jk'}+r_{kk'} ~.
\end{equation}
In this case, assuming 
that the neutrino effective Dirac masses are much smaller than the effective 
right-handed neutrino Majorana masses, the effective $3\times 3$ active neutrino 
mass matrix $m_{\nu}$ is  proportional to, schematically, 
\begin{eqnarray}
m_{ij} = \frac{v^2}{\Lambda}\left(c_\ell^{ij}\epsilon^{|q_{ij}|}  \right.
& + & \sum_{k,k'}\lambda_{\nu}^{ik} (c_n \epsilon^{|r|})^{-1}_{kk'} (\lambda_{\nu}^{\top})^{k'j}
 \nonumber \\ [0.3em]
&\times &\left.\epsilon^{|p_{ik}|+|p_{jk'}|}\right) ~,
\end{eqnarray}
where $c_{n} \epsilon^{|r|}$ is a short-hand notation for a matrix whose $(k,k^{\prime})$ entry is $c_{n}^{kk^\prime} \epsilon^{|r_{kk^{\prime}}|}$.  

As opposed to usual seesaw models where the smallness of the neutrino mass is 
due to $\Lambda\gg \vev$, in our theory the neutrinos are naturally light 
even when $\Lambda\lesssim 10$~TeV, provided $\epsilon$ is small enough, or 
its exponents are large enough.
In particular, if there are no integer values of $q_{ij}$ and $r_{kk^\prime}$, 
then the neutrinos are purely Dirac fermions, and their masses are naturally small 
if  $p_{ik}$ are large integers.
Note also that for large values of $r_{kk'}$ there are 
light -- sometimes very light -- mostly-sterile neutrinos.
One should keep in mind, though,  
that if the exponents are too large, then the operators shown 
in Eq.~(\ref{general_mass}) would have a high mass dimension, and embedding 
them in a renormalizable model would require an extended set of heavy vectorlike 
fermions 
\cite{Froggatt:1978nt}.

The choice of $U(1)_{\nu}$ charges also allows one to probe natural explanations 
for the neutrino mass hierarchy, and the pattern of the leptonic mixing matrix. 
The reason for this is that the $U(1)_{\nu}$ charges for the leptons can be family 
dependent, in which case $U(1)_{\nu}$ operates as a family symmetry.    
In this particular approach to flavor, it is natural to obtain a hierarchy among the
 elements of the neutrino (and charged lepton) mass matrices (e.g. $M_{11}\gg M_{12}$, etc). 
 On the other hand, this approach does not explain the relationship between
different entries in the mass matrices that have, naively, 
the same order of magnitude. This is because all the dimensionless coefficients 
$\lambda_{\nu}$, $c_{\ell}$, and $c_{n}$ are undetermined free parameters in our framework.

\subsection{Charged leptons and flavor-changing neutral currents}
\label{sec:formalism-fcnc}

While we have imposed constraints on the charged-lepton $U(1)_{\nu}$ charges to guarantee that renormalizable diagonal charged-lepton Yukawa interactions are allowed by gauge invariance, the presence of off-diagonal Yukawa interactions will depend on specific choices for $z_{\ell_i}$ and $z_{e_i}$. If these are not allowed by gauge invariance at the renormalizable level, there remains the possibility that, similar to the neutrino mass operators discussed in the previous subsection,  
higher-dimensional operators will generate them:
\be
\sum_{ij} \lambda_e^{ij} \left(\frac{g_\phi}{\Lambda} \phi \right)^{s_{ij}} 
\overline{\ell}_{L}^{i} e_{R}^{j}H  + {\rm H.c.} ~.
\label{cl_mass}
\ee 
After the constraints discussed earlier are taken into account, the exponents
$s_{ij}$ are given by
\begin{equation}
s_{ij}=z_{\ell_i}-z_{\ell_j} ~.
\end{equation} 
As with the neutrinos, it is understood that only terms corresponding to integer values of $s_{ij}$ are present.
Note that $s_{ij}\equiv0$ for $i=j$, so Eq.~(\ref{cl_mass}) includes Eq.~(\ref{cc_mass}). 

Off-diagonal charged lepton Yukawa interactions will lead to tree-level, charged-lepton flavor violating couplings of the $Z'$ gauge boson if the charged-lepton $U(1)_{\nu}$ charges are flavor dependent ($z_{\ell_i}\neq z_{\ell_j}$ or $z_{e_i}\neq z_{e_j}$ for some $i\neq j$). These, in turn, will mediate charged-lepton flavor violating processes such as $\mu^+\to e^+e^-e^+$, $\mu \!\rightarrow e$ conversion in nuclei, $\tau\to \ell\ell'\ell''$, and $\tau\to \ell+$hadron(s). The rates of such processes will be, naively, $\Gamma\propto(\epsilon^{|s_{ij}|}/\langle\phi\rangle^2)^2$. Current limits on charged-lepton flavor violation require $\epsilon^{|s_{ij}|}\ll 1$ or $\langle \phi \rangle\gg v$. Note that if $U(1)_{\nu}$ charges are such that there are no integer 
$s_{ij}$ with $i\neq j$, then $Z'$ exchange will not mediate charged-lepton flavor violating processes at tree level, regardless of the relative values of the the charged-lepton $U(1)_{\nu}$ charges. The reason for this is the fact that both the $Z'$ coupling and the Yukawa coupling charged-lepton eigenbases are the same ({\it i.e.}, both matrices are diagonal) in this case. 

Similar to what happens with the charged leptons, off-diagonal neutral lepton interactions will also lead to flavor-changing processes in the neutrino sector. Unlike processes involving charged leptons, however, neutrino flavor-changing neutral currents are not very severely constrained \cite{neutrino_fcnc}. This means that these are not likely 
to be cause for concern as long as $\langle \phi\rangle$ is only slightly larger than the electroweak scale, regardless of the relative strength of the off-diagonal neutrino--$Z'$ coupling. On the other hand, phenomenologically interesting choices for the lepton $U(1)_{\nu}$ charges often predict large neutrino flavor-changing processes, which will be significantly better constrained in the next round of neutrino oscillation experiments \cite{neutrino_fcnc_future}.

\setcounter{equation}{0}\setcounter{footnote}{0}
\section{Leptocratic Model}
\label{sec:model}

In order to construct a viable theory for physics beyond the standard model 
using the framework presented in Sec.~\ref{sec:formalism}, one has to identify a 
set of commensurate charges for the right-handed neutrinos which cancel the 
$[U(1)_\nu]^3$ anomaly, as shown in Eq.~(\ref{cubic}).
This is a highly nontrivial problem: cubic equations do not have 
integer solutions except for very special cases.
In this section we show that in the case of $N=3$ right-handed neutrinos
the cubic equation can be solved for arbitrary lepton charges consistent 
with the restrictions imposed in Sec.~\ref{sec:chargedmasses}. 

We have mentioned at the end of Sec.~\ref{sec:chargedmasses} that altogether there are 
two quark charges, two lepton-doublet charges and $N-1$ right-handed neutrino
charges that remain independent once all standard model mass terms are allowed, 
and all anomaly cancellation conditions other than the $[U(1)_\nu]^3$ are imposed.
The cubic equation Eq.~(\ref{cubic}) in these six rational variables (for $N=3$) appears 
daunting at first sight. However, we found a simple parametrization 
of the charges that greatly simplifies its solution.
The lepton-doublet charges may be written in terms of $z_q$ and two
rational numbers, $a$ and $a^\prime$:
\bear
&& z_{\ell_1} \equiv - 3 z_q -2a ~, 
\nonumber \\ [0.2em]
&& z_{\ell_2}  \equiv - 3 z_q + a + a^\prime ~, 
\nonumber \\ [0.2em]
&& z_{\ell_3} = - 3 z_q + a - a^\prime ~, 
\label{leptocratic-1}
\eear
so that the second sum rule of Eq.~(\ref{eq:zl3}) is automatically satisfied.
The charges of the right-handed quarks are conveniently parametrized as
\bear
&& z_u \equiv 4 z_q - \frac{c}{2} ~~,
\nonumber \\ [0.3em]
&& z_d = -2 z_q + \frac{c}{2}  ~~,
\eear
where $c$ is a rational number.
The Higgs doublet has $U(1)_\nu$ charge
\be
z_{\tiny H} = 3 z_q - \frac{c}{2}   ~~.
\ee
The $U(1)_\nu$ charges of the $SU(2)_L$-singlet standard model leptons are 
then fixed by requiring renormalizable mass terms for the charged leptons
[see Eqs.~(\ref{eq:e2e3}) and (\ref{eq:e1})]:
\bear
&& z_{e_1} \equiv -6 z_q + \frac{c}{2} - 2a ~, 
\nonumber \\ [0.2em]
&& z_{e_2}  \equiv - 6 z_q  + \frac{c}{2} + a + a^\prime ~, 
\nonumber \\ [0.2em]
&& z_{e_3} = - 6 z_q  + \frac{c}{2} + a - a^\prime ~.
\eear
Finally, the right-handed neutrinos have charges
\bear
&& z_{n_1} \equiv -\frac{c}{2} - 2b ~, 
\nonumber \\ [0.2em]
&& z_{n_2}  \equiv -\frac{c}{2} + b + b^\prime ~, 
\nonumber \\ [0.2em]
&& z_{n_3} = -\frac{c}{2} + b - b^\prime ~, 
\label{leptocratic-5}
\eear
which automatically satisfy the sum rule given in Eq.~(\ref{linear}) for 
any rational numbers  $b$ and $b^\prime$.
With this parametrization, 
all the terms involving $z_q$ from the cubic equation drop out, 
such that the $[U(1)_\nu]^3$ anomaly cancellation condition (\ref{cubic})
takes a rather simple form:
\be 
\frac{c}{2} \left(3a^2 + a^{\prime 2} - 3b^2 - b^{\prime 2} \right)
= - a \left(a^2 - a^{\prime 2}\right) +  b \left(b^2 - b^{\prime 2}\right)  ~.
\label{LOM-cubic}
\ee
Remarkably, only terms linear in $c$ or independent of $c$ are present,
so that for any rational numbers $a,a^\prime, b$ and $b^\prime$ there is a 
rational solution for $c$. 

Given that only the lepton charges in this model are allowed to be generation dependent,
we will refer to the charge assignment described by Eqs.~(\ref{leptocratic-1})-(\ref{LOM-cubic}) 
as the ``Leptocratic Model''. 
 
The exponents that determine the orders of magnitude of the various Dirac neutrino 
masses are given by 
\bear 
&& \hspace*{-1.3em} p= \left(\begin{array}{ccc}
-2(a-b) \; & \;\; - \left(2a + b + b^\prime\right)  \; & \;\; - \left(2a + b - b^\prime\right) \\ [0.2em]
a + a^\prime + 2b  \;\; & \; a + a^\prime - b - b^\prime   \; & \;\;   a + a^\prime - b + b^\prime  \\ [0.2em]
a - a^\prime + 2b  \;\; & \; a - a^\prime - b - b^\prime   \; & \;\;   a - a^\prime - b + b^\prime  \\ [0.2em]
\end{array}\right)~. \nonumber \\ [0.18em] &&
\eear
The orders of magnitude of the left-handed
Majorana neutrino masses are determined by the following exponents:
\bear 
&& \hspace*{-.3em} q = \left(\begin{array}{ccc}
c + 4a           \; & \;\; c + a - a^\prime     \; & \;\; c + a + a^\prime  \\ [0.2em]
c + a - a^\prime \; & \;\; c - 2\left(a+a^\prime\right) \; & \;\;  c - 2a  \\ [0.2em]
c + a + a^\prime \; & \;\; c - 2a  \; & \;\; c - 2\left(a-a^\prime\right) \\ [0.2em]
\end{array}\right) ~~. \nonumber \\ [0.18em] &&
\eear
For right-handed Majorana neutrino masses, the exponents are given by
\bear
&& \hspace*{-.3em}
r = \left(\begin{array}{ccc}
c + 4b           \; & \;\; c + b - b^\prime          \; & \;\; c + b + b^\prime  \\ [0.2em]
c + b - b^\prime \; & \;\; c - 2\left(b+b^\prime\right) \; & \;\;  c - 2b  \\ [0.2em]
c + b + b^\prime \; & \;\; c - 2b  \; & \;\; c - 2\left(b-b^\prime\right) \\ [0.2em]
\end{array}\right) ~~.  \nonumber \\ [0.18em] &&
\eear
The exponents in the charged lepton mass matrix are
\bear
&& \hspace*{-.3em}
s=\left(\begin{array}{ccc}
0 & \; -3a-a^{\prime} \; & \; -3a + a^{\prime} \\ [0.2em]
3a+a^{\prime} & 0 & 2a^{\prime} \\ [0.2em]
3a-a^{\prime} & -2a^{\prime} & 0 \\ [0.2em]
\end{array}\right) ~~.
\nonumber \\ [0.18em] &&
\eear

If the factor multiplying $c$ in Eq.~(\ref{LOM-cubic}) vanishes, then one can show that some of the 
$p_{ij}$ exponents vanish. Given that we want to avoid dimension-4 Dirac neutrino mass
terms, we assume 
\be
3a^2 + a^{\prime 2} \neq 3b^2 + b^{\prime 2} ~,
\ee
which implies 
\be
c = - 2 \, \frac{a\left(a^2 - a^{\prime 2}\right) - b\left(b^2 - b^{\prime 2}\right)}
{3 a^2 + a^{\prime 2} - 3 b^2 - b^{\prime 2}} ~.
\label{cab}
\ee

The orders of magnitude of various neutrino mass terms are set by the exponents
$p_{ij}$, $q_{ij}$ and $r_{ij}$, where it is understood that nonzero entries exist only if 
the corresponding exponent is an integer.
The parameter space that includes these exponents has four arbitrary rational parameters
$a, a^\prime, b, b^\prime$. For most choices of these parameters, the neutrino 
phenomenology is not consistent with the observations. However, there are cases
where all phenomenological constraints are satisfied. We do not attempt here to search 
for all such cases, but rather give some interesting examples.


\subsection{``Orwellian'' Leptocratic Model}
\label{sec:Orwellian}

\begin{table}[t]
\renewcommand{\arraystretch}{1.5}
\begin{tabular}{|c|c|}\hline 
\ field \ & \ $U(1)_\nu$ charge \ \\  [0.03em] \hline\hline
$q_L$ & \ $z_q$ \ \\ [0.1em]
$u_R$ & \ $\displaystyle 4z_q + \frac{b}{3}$ \ \\ [0.4em]
$d_R$ & \ $\displaystyle -2z_q - \frac{b}{3}$ \ \\  [0.4em] \hline
$\ell_L$ & \ $-3z_q$ \ \\ 
$e_R$ & \ $\displaystyle -6z_q - \frac{b}{3}$ \ \\  [0.3em]
$n_R^1$ & \ $\displaystyle - \frac{5b}{3}$ \ \\  [0.5em]
\ $n_R^2, n_R^3$ \ & \ $\displaystyle \frac{4b}{3}$ \ \\  [0.4em] \hline
$H$ & $\displaystyle 3z_q + \frac{b}{3}$ \ \\
$\phi$ & $+1$ \\ \hline
\end{tabular}
\medskip \\
\vspace{0.2in}
\caption{\small $U(1)_\nu$ charges of the quarks, leptons and scalars,
for the Leptocratic Model with $a = a^\prime = b^\prime = 0$. }
\end{table}

Let us first consider the case where the $U(1)_\nu$ charges of the 
standard model fermions are generation-independent, 
and the 
charges of the second and third right-handed neutrinos are 
equal but different from the charge of the first right-handed neutrino:
\be
a = a^\prime = b^\prime = 0 ~~.
\ee
In this case, there are no charged-lepton flavor violating processes mediated by $Z'$ exchange at the tree level.
The $U(1)_\nu$ charges are shown in Table I.
We will refer to this particular type of Leptocratic Model
as Orwellian \cite{orwell}, given that one of the right-handed neutrinos
has a $U(1)_\nu$ charge different from the other two. Note that we do not consider 
the case $b=0$ because that would 
allow Dirac masses from dimension-4 operators. 
Under these circumstances, Eq.~(\ref{cab}) implies 
\be
c = -\frac{2}{3} b ~.
\ee

Depending on the values of $b$, there are three viable cases with different 
phenomenology.
First, if $b$ is an integer but not a multiple of 3, then 
all Dirac mass terms are allowed,
 \begin{equation}
M_D = v\,
\epsilon^{|b|}
\left(\begin{array}{ccc}
\lambda_\nu^{11} \, \epsilon^{|b|} & \;\;\; \lambda_\nu^{12} & \;\;\;  \lambda_\nu^{13} \\ [0.2em]
\lambda_\nu^{21} \, \epsilon^{|b|} & \;\;\; \lambda_\nu^{22} & \;\;\; \lambda_\nu^{23} \\ [0.2em]
\lambda_\nu^{31} \, \epsilon^{|b|} & \;\;\; \lambda_\nu^{32} & \;\;\; \lambda_\nu^{33} 
\end{array}\right)~~,
\label{OLM-dirac}
\end{equation}
while all Majorana masses are forbidden by gauge invariance.
Assuming that all Yukawa-like couplings $\lambda_\nu^{ij}$ are of the same order of magnitude,
the above Dirac mass matrix provides a good fit to all existing neutrino data (except for 
those from LSND). This is an interesting case because it shows that the neutrinos may be 
Dirac fermions with naturally small masses. For example, if all $\lambda_\nu^{ij}$ are in the range of 
 $10^{-5}-10^{-4}$ ({\it i.e.}, somewhat larger than the electron Yukawa coupling), 
and $\epsilon \sim 10^{-4}$, then for $b= \pm 2$ one can easily obtain a neutrino mass spectrum with
normal hierarchy:
$m_{\nu_3} \sim 0.05$ eV, $m_{\nu_2} \sim 0.01$ eV (to accommodate the 
atmospheric and solar mass-squared differences), and $m_{\nu_1}  \sim 10^{-9}$ eV.\footnote{Another possibility is to choose $\lambda_{\nu}^{ij}\sim1$, $\epsilon\sim 0.1$ and $|b|=13$. In this case, the lightest neutrino mass is $10^{-13}$ times the mass of the two heaviest states.}  The 
extra factor of $\epsilon^{|b|}$ in the first column of the mass matrix $M_D$ implies that one of the 
neutrinos is many orders of magnitude lighter than the other two. Lepton mixing, on the other hand,
is anarchical, in good agreement with the current oscillation data \cite{anarchy}.  
This particular approach to flavor generically leads to a normal neutrino mass
hierarchy if the neutrinos are Dirac fermions. 
The reason for this -- as discussed in Sec.~\ref{sec:formalism} -- is that while the order of magnitude of the ratios of the different
entries in the mass matrices are predicted, there are still ``order one'' coefficients that lead to 
order one differences among matrix elements that correspond to the same suppression factor. 
This means that while we can choose charges so that two neutrino masses are of the same order of 
magnitude, there is no ingredient within the formalism pursued here that guarantees that these two
masses are quasi-degenerate (which is required if the inverted neutrino mass hierarchy is realized).
A very similar situation is encountered in ``fat-brane'' models \cite{fat-brane}.

The second viable case is that where  $b$ is a multiple of 3.
As a result,  $c$ is an integer so that all Majorana and Dirac mass terms are allowed. 
The left-handed Majorana masses are given by 
 \begin{equation}
M_L = \frac{v^2}{\Lambda} \,
\epsilon^{2|b|/3} \; c_\ell ~~,
\label{OLM-LH}
\end{equation}
the Dirac masses 
are given by Eq.~(\ref{OLM-dirac}), and the 
right-handed Majorana masses are given by
\bear 
&& \hspace*{-.3em}
M_{R}= \Lambda \,
\epsilon^{|b|/3} 
\left(\begin{array}{ccc}
c_n^{11} \epsilon^{3|b|} & \;\;\; c_n^{12} & \;\;\; c_n^{13} \\ [0.2em]
c_n^{12}  & \;\;\; c_n^{22} \epsilon^{7|b|/3} & \;\;\; c_n^{23} \epsilon^{7|b|/3} \\ [0.2em]
c_n^{13}  & \;\;\; c_n^{23}  \epsilon^{7|b|/3}& \;\;\; c_n^{33}  \epsilon^{7|b|/3}
\end{array}\right) ~~. \nonumber \\ [0.18em] &&
\eear
Assuming that all coefficients of the higher-dimensional operators in Eq.~(\ref{general_mass}),
$c_\ell^{ij}$, $c_n^{ij}$, $\lambda_\nu^{ij}$, are of the same order of magnitude, 
then for  $\Lambda \sim O(1$ TeV) and $\epsilon \ll 1$ 
we find that the left-handed Majorana masses are larger than the 
Dirac ones, two of the eigenvalues of $M_{R}$ are substantially larger, and the third one is 
very small. 
Therefore the three active neutrinos are Majorana fermions with masses of approximately
the same order of magnitude. 
The solar and atmospheric oscillation data are well accommodated for a significant range 
of coefficients $c_\ell^{ij}$, as established for ``anarchical''
left-handed Majorana masses \cite{anarchy}.
Besides the three active neutrinos there are two heavy sterile neutrinos
which have small mixing with the active ones, and a light sterile neutrino
whose mixing with the active neutrinos may be phenomenologically relevant. 
We relegate the detailed derivation of their mixing to the Appendix, and summarize 
the results below. 

The lightest sterile neutrino has a mass given by an ``inverted'' seesaw mechanism:
\bear
m_{\nu_4} 
& = & \mathcal{O}\biggl( \frac{(\lambda_\nu^{ij})^{2}}{c_{\ell}^{ij}}\biggr) \Lambda \, \epsilon^{4|b|/3}~,
\eear 
while the masses of the two heavier sterile neutrinos are
\begin{equation}
m_{\nu_{5,6}} = \mathcal{O}( c_{n}^{ij}) \Lambda \epsilon^{|b|/3} ~.
\end{equation}
If one assumes all coefficients from Eq.~(\ref{general_mass})
to be of the order
of $10^{-5}$ and $\epsilon \sim 10^{-4}$, for $b= \pm 3$ the neutrino 
spectrum may naturally be given by 
$m_{\nu_3} \sim 0.05$ eV and $m_{\nu_2} \sim m_{\nu_1} \sim 0.01$ eV. The 
two heavy sterile neutrinos have masses of order 1 keV, and the remaining 
sterile neutrino has a mass of order $10^{-9}$ eV. 
The mixing between the active neutrinos and lightest sterile neutrino is
\begin{equation}
\Theta_{\rm active-light} \sim \epsilon \, \frac{\Lambda}{v} \sim  10^{-3} \; , 
\end{equation}
and the mixing between the active neutrinos and the two heavy sterile neutrinos is
\begin{equation}
\Theta_{\rm active-heavy} \sim \epsilon^{2} \frac{v}{\Lambda} \sim 10^{-9} \; .
\end{equation}
While very weakly coupled, these sterile neutrinos can lead to interesting phenomenology. 
The lightest of the sterile neutrinos, for example, can lead to observable consequences in 
precision measurements of low energy solar neutrino oscillations \cite{deHolanda:2003tx}. 
Other neutrino oscillation experiments might be sensitive to these quasi-sterile neutrinos 
if a less constrained parameter space for $\lambda_{\nu}, \; c_{n}, \; c_{\ell}, \; \epsilon$ and $\Lambda$ 
is considered. 

Another possibility is to have all neutrino Yukawa couplings, $\lambda_{\nu}$, $c_{\ell}$ and $c_{n}$, of order $\mathcal{O}(1)$, while having $\epsilon \sim 0.1$ for $b=\pm 18$. The neutrino mass spectrum is then given by, $m_{\nu_{3}} \sim 0.05$ eV, $m_{\nu_{1,2}} \sim 0.01$ eV, while the two heavy sterile ones have masses around 1~MeV and the lightest sterile neutrino weighs about $10^{-12}$ eV. The mixing between the active and heavy sterile neutrinos is
\begin{equation}
\Theta_{\rm active-heavy} \sim \epsilon^{12} \frac{v}{\Lambda} \sim 10^{-13} \; ,
\end{equation}
while the active-light sterile mixing is
\begin{equation}
\Theta_{\rm active-light} \sim \epsilon^{6} \frac{\Lambda}{v} \sim 10^{-5} \; .
\end{equation}

The third and last viable case is that where $b$ is an odd multiple of 3/2.
The left-handed Majorana masses are then all allowed as in the previous case, being 
given by Eq.~(\ref{OLM-LH}).
The Dirac mass terms associated with $\lambda^{ij}_{\nu}$  are strictly 
forbidden by gauge invariance for $j=2$ and $j=3$, and equal to
$\lambda^{i1}_{\nu} v\epsilon^{2|b|}$ for $j=1$ and any $i=1,2,3$.
The right-handed Majorana masses are given by
 \begin{equation}
M_{R}= \Lambda \,
\epsilon^{8|b|/3} 
\left(\begin{array}{ccc}
c_n^{11} \epsilon^{2|b|/3} & \;\;\; 0 & \;\;\;  0 \\ [0.2em]
0  & \;\;\; c_n^{22} & \;\;\; c_n^{23} \\ [0.2em]
0  & \;\;\; c_n^{23} & \;\;\; c_n^{33} 
\end{array}\right)
~~.
\end{equation}
If all $c_\ell^{ij}$, $c_n^{ij}$ and $\lambda_\nu^{ij}$ are of the same order of magnitude, 
then for  $\Lambda \sim O(1$ TeV) and $\epsilon \ll 1$
the left-handed Majorana masses dominate,
so again the anarchical solution works well.
In this case the right-handed neutrinos are ultra-light fermions 
that couple to the standard model only via the $Z^\prime$ boson.
For example, if all coefficients are of the order
of $10^{-5}$ and $\epsilon \sim 10^{-2}$, we find that for $b= \pm 9/2$ 
the three active neutrinos have masses in the $0.01 - 0.1$ eV range,
two sterile neutrinos have masses of about $10^{-17}$ eV, and the 
remaining sterile neutrino is even lighter, with a mass of $10^{-23}$ eV. 
There is no mixing between the active neutrinos and the heavy sterile ones, 
while the mixing between the active and the lightest sterile neutrinos is of order  
$10^{-11}$. Another viable case is to have all coefficients of order $1$ 
and $\epsilon \sim 0.1$. Active neutrino masses $\sim 0.01 - 0.1$ eV 
can also be accommodated with $b = \pm 33/2$. In this case, the heavy and light sterile neutrinos have masses 
$\sim 10^{-32}$ and $\sim 10^{-43}$ eV, respectively, while the mixing between the active and 
lightest sterile neutrinos is of order $10^{-21}$.

\subsection{$2+1$ Leptocratic Model}

One can further relax the assumption of having generation independent $U(1)_{\nu}$ charges for the lepton doublets by allowing a non-zero value for the parameter $a$. A viable mixing pattern for the neutrinos can arise if the $U(1)_{\nu}$ charges for the second and third generations are identical for the lepton doublets and for the right-handed neutrinos, {\it i.e.} $a^{\prime} = b^{\prime}=0$, while  $a$ and $b$ are non-zero. The Dirac neutrino mass matrix then becomes,
\begin{equation}
M_{D} = v \left(
\begin{array}{ccc}
\lambda_{\nu}^{11}  \epsilon^{|-2 (a-b)|} &  \lambda_{\nu}^{12} \epsilon^{|-2a-b|}  & \lambda_{\nu}^{13} \epsilon^{|-2a-b|}
\\
\lambda_{\nu}^{21} \epsilon^{|a+2b|} & \lambda_{\nu}^{22} \epsilon^{|a-b|} & \lambda_{\nu}^{23} \epsilon^{|a-b|}
\\
\lambda_{\nu}^{31} \epsilon^{|a+2b|} & \lambda_{\nu}^{32}  \epsilon^{|a-b|} & \lambda_{\nu}^{33} \epsilon^{|a-b|} 
\end{array}\right) \; .
\end{equation}
In the Orwellian Leptocratic case with universal charges for the lepton doublets, the coefficients $\lambda_{\nu}^{ij}$ play a determining role when it comes to ``explaining'' the large mixing angles and the neutrino  mass-squared splittings required by the atmospheric and the solar neutrino data. In the present case with non-universal charges for lepton doublets, however, the structure that is needed for the bi-large mixing pattern and the mass splittings is already built-in due to the hierarchy among the exponents of $\epsilon$ in the entries of the mass matrices. 
The right-handed Majorana mass matrix in this case is given by,
\begin{equation}
M_{R} =  \Lambda \left(\begin{array}{ccc}
c_{n}^{11} \epsilon^{|c+4b|} & c_{n}^{12} \epsilon^{|c+b|} & c_{n}^{13} \epsilon^{|c+b|}
\\
c_{n}^{21} \epsilon^{|c+b|} & c_{n}^{22} \epsilon^{|c-2b|} & c_{n}^{23} \epsilon^{|c-2b|}
\\
c_{n}^{31} \epsilon^{|c+b|} & c_{n}^{32} \epsilon^{|c-2b|} & c_{n}^{33} \epsilon^{|c-2b|}
\end{array}\right) \; , 
\end{equation}
and the left-handed Majorana mass matrix is,
\begin{equation}
M_{L} = \frac{v^{2}}{\Lambda} \left(\begin{array}{ccc}
c_{\ell}^{11} \epsilon^{|c|} & c_{\ell}^{12} \epsilon^{|c+a|} & c_{\ell}^{13} \epsilon^{|c+a|}\\
c_{\ell}^{21} \epsilon^{|c + a|} & c_{\ell}^{22} \epsilon^{|c-2a|} & c_{\ell}^{23} \epsilon^{|c-2a|}\\
c_{\ell}^{31} \epsilon^{|c+a|} & c_{\ell}^{32} \epsilon^{|c-2a|} & c_{\ell}^{33} \epsilon^{|c-2a|}
\end{array}\right) \; ,
\end{equation}
where the parameter $c$ is now given in terms of $a$ and $b$ as,
\begin{equation}
c 
= -\frac{2}{3} \left(\frac{a^{2} + ab + b^{2}}{a+b}\right) \; .
\end{equation}

If $a$ and $b$ are chosen such that the parameter $c$ is non-integer, then 
the Dirac mass matrix is allowed while both the left-handed and right-handed Majorana mass matrices are forbidden by gauge invariance.  With $\alpha = a - b$, the neutrino Dirac mass matrix  can be rewritten as,
\begin{equation}
M_{D} = v \left(\begin{array}{ccc}
\lambda_{\nu}^{11} \epsilon^{|2\alpha|} & \lambda_{\nu}^{12} \epsilon^{|2\alpha+3b|} & \lambda_{\nu}^{13} \epsilon^{|2\alpha+3b|}
\\
\lambda_{\nu}^{21} \epsilon^{|2\alpha-3a|} & \lambda_{\nu}^{22} \epsilon^{|\alpha|} & \lambda_{\nu}^{23} \epsilon^{|\alpha|} 
\\
\lambda_{\nu}^{31} \epsilon^{|2\alpha-3a|} & \lambda_{\nu}^{32} \epsilon^{|\alpha|} & \lambda_{\nu}^{33} \epsilon^{|\alpha|}
\end{array}\right) \; .
\end{equation}
The ratio $\Delta m_{sol}^{2}/ \Delta m_{atm}^{2} \sim 10^{-2}$ indicates that the ratio of the $(1,3)$ to the $(2,3)$ elements of the neutrino Dirac mass matrix $M_{D}$ is of order $\sim 0.1$. Because only exponents that are integers are allowed in the mass matrix, the parameter $\epsilon =  g_{\phi} \left<\phi \right > / \Lambda$ cannot be smaller than $\mathcal{O}(0.1)$, given that the smallest possible value for the difference between the exponents $| 2\alpha+3b  |$ and $| \alpha  |$ is $1$.   
One viable example is to have $a=25/3$, $b=-11/3$ and $\epsilon \sim 0.1$ with all Yukawa 
couplings $\lambda_{\nu}^{ij}$ of order one, leading to the following Dirac mass matrix,
\begin{equation}
M_{D} = v \, \epsilon^{12} \left(\begin{array}{ccc}
\mathcal{O}( \epsilon^{12}) & \mathcal{O}( \epsilon) & \mathcal{O}( \epsilon)
\\
\mathcal{O}( \epsilon^{13}) & \mathcal{O}(1) & \mathcal{O}( 1)
\\
\mathcal{O}( \epsilon^{13}) & \mathcal{O}( 1)  & \mathcal{O}( 1)
\end{array}\right) \; ,
\end{equation}
which naturally gives rise to a normal mass hierarchy as well as bi-large mixing among neutrinos of the Dirac type. The smallness of the neutrino masses is a consequence of the large exponent $|\alpha|$ if $\Lambda,\langle\phi\rangle\sim \mathcal{O}(\mbox{1 TeV})$.
As mentioned in Sec.~\ref{sec:formalism}, such high dimensional operators pose a 
challenge for constructing elegant underlying theories.

In this case, because $z_{l_1}\neq z_{l_2}=z_{l_3}$, we must check whether $Z'$ exchange 
mediates very fast charged-lepton flavor violating processes. 
The (1,2) and (1,3) elements in the charged lepton mass matrix are proportional to $\epsilon^{|s_{12}|}$ and $\epsilon^{|s_{13}|}$ (see Eq.~(\ref{cl_mass})), assuming the $s_{ij}$ are integers. While nonzero, $s_{12}$ and $s_{13}$ turn out to be huge: $|s_{12}|=|s_{13}|=3a=25$, so that charged lepton flavor violating phenomena are highly suppressed (by a factor $10^{-25}$). Because  $z_{l_2}=z_{l_3}$, off-diagonal (2,3) elements will  not lead to tree-level flavor changing effects.


\setcounter{equation}{0}\setcounter{footnote}{0}
\section{Other models}
\label{sec:cases}

The Leptocratic Model described in Sec.~\ref{sec:model} is based on several assumptions 
that could in principle be relaxed. 
For example, if the scalar sector is enlarged to include two or more 
$SU(2)_L$-singlet scalars carrying different $U(1)_\nu$ charges, then 
several elements of the neutrino mass matrix may get larger contributions from 
gauge invariant operators involving the new $\phi$ fields. The same happens 
in the presence of two or more $SU(2)_L$-doublet scalars with carefully chosen 
$U(1)_\nu$ charges.

Another assumption of the Leptocratic Model is that the electrically-charged leptons
have renormalizable mass terms. If instead the flavor-diagonal mass terms 
of the electron, the muon and possibly the tau are generated by 
higher-dimensional operators as in Eq.~(\ref{cl_mass}), then Eqs.~(\ref{eq:e2e3}) and (\ref{eq:e1}),
which determine the $U(1)_\nu$ charges of the $SU(2)_L$-singlet leptons, are modified as follows:
\be
z_{e_i} = z_{\ell_i} - z_u + z_q - s_{ii} ~.
\ee
The integers $s_{ii}$, $i=1, \; 2, \; 3$, are constrained by the $[U(1)_Y]^2 U(1)_\nu$ anomaly cancellation 
to satisfy
\be
s_{33} = -s_{11} - s_{22} ~,
\ee
so that two new discrete parameters are introduced to the theory. 
The advantage of this approach is that the hierarchy of lepton masses may be explained without a 
hierarchy among dimensionless parameters. 
The disadvantage is that the ratio of neutrino and
electron masses suggests that the neutrino masses are generated by very high-dimension 
operators, which may be difficult to realize in an underlying theory that is well-behaved in the 
ultra-violet. 

In the remainder of this section we explore some cases where the 
number of right-handed neutrinos is different from three, without relaxing any 
other assumption of the Leptocratic Model.
We have shown in Sec.~\ref{sec:model} that the cubic equation describing 
$[U(1)_\nu]^3$ anomaly cancellation has a general solution with commensurate charges
when there are $N=3$ right-handed neutrinos. We have not found general
solutions of this type for $N=2$ or $N\ge 4$. To understand the problem, 
consider the case $N=2$. Using the parametrization of charges given in 
Eqs.~(\ref{leptocratic-1})-(\ref{leptocratic-5}),
we can analyze the anomaly cancellation conditions in this case by 
imposing that one of the right-handed neutrino charges vanishes in the $N=3$ case.
For example, $z_{n_1} = 0$ implies $c = -4b$, so that 
the cubic equation in the $N=2$ case takes the form 
\be
5 b^3 + b \left(3 b^{\prime 2} -6a^2 -2 a^{\prime 2} \right) 
= - a \left(a^2 - a^{\prime 2}\right) ~.
\label{cubic-n2}
\ee
Different from the cubic equation for the $N=3$ case [see Eq.~(\ref{LOM-cubic})], 
this equation is not linear in any of its four variables, so that a general 
solution with commensurate charges is hard to find. 
However, such solutions can be easily identified in certain special cases. 

For example, the $N=2$ case with all standard model fermions having 
generation independent charges
implies $a = a^\prime = 0$, and we find that the cubic equation Eq.~(\ref{cubic-n2})
is satisfied for $b=0$ and arbitrary $b^\prime$. It follows that the two right-handed 
neutrinos form a vectorlike fermion ($z_{n_2} = - z_{n_3}\equiv z$), and the 
$U(1)_{\nu}$ charges of the standard model fermions are proportional to their 
hypercharges \cite{Appelquist:2002mw}.
Schematically, the $5\times 5$ neutrino mass matrix has the following form:
\begin{equation}
M_{\nu}=\left(\begin{array}{c|cc} \displaystyle
\frac{v^{2}}{\Lambda} \left( c_\ell\right)_{3\times3}
& v \epsilon^{|z|} \left( \lambda_\nu\right)_{3\times 1}
& v \epsilon^{|z|} \left( \lambda_\nu\right)_{3\times 1}
\\[1em]  \hline 
&& \\ [-1em] 
& c_n^{11} \Lambda\epsilon^{|2z|} &  c_n^{12} \Lambda
\\ [1em]
& &  c_n^{22} \Lambda\epsilon^{|2z|}
\end{array}\right).
\end{equation}
This is a symmetric matrix, and we have shown only the upper off-diagonal
elements. It is also understood that the terms that include powers of $\epsilon$ 
vanish if the exponent is non-integer.
Given that all left-handed Majorana mass terms arise from gauge invariant dimension-5
operators, the smallness of the neutrino masses suggests $\Lambda\gg v$. Thus, the two right-handed 
neutrinos pair up to form a mostly pseudo-Dirac fermion of large mass, of order $\Lambda$. Their mixing with 
the active states is very small if $z$ is an integer, and it vanishes otherwise.
The light neutrino sector is identical to the one in models with a 
standard high-energy seesaw mechanism.


For $N=1$,  
all anomaly cancellation conditions can be solved analytically, 
such that all $U(1)_{\nu}$ charges can be expressed in terms of three 
independent charges, chosen here to be $z_q$, $z_{\ell_1}$, $z_{\ell_2}$.
There is one solution where the right-handed neutrino and quark charges 
are given by 
\begin{eqnarray}
z_{n} & = & \frac{3}{2}\left( 3z_{q} + z_{\ell_1} \right)~,\nonumber\\[0.4em]
z_{u} & = & \frac{1}{2}\left(11 z_{q} + z_{\ell_1}\right) ~,\nonumber\\ [0.4em]
z_{d} & = & -\frac{1}{2}\left(7 z_{q} + z_{\ell_1}\right) ~.
\label{eq:one-n}
\end{eqnarray}  
The electrically charged $SU(2)_L$-singlet leptons then have the following $U(1)_{\nu}$ charges:
\begin{eqnarray}
z_{e_1}  & = &   -\frac{1}{2}\left(9z_{q} - z_{\ell_1}\right) ~,\nonumber\\ [0.4em]
z_{e_2} & = &  -\frac{1}{2}\left(9z_{q} + z_{\ell_1}\right) +  z_{\ell_2} ~,\nonumber\\ [0.4em]
z_{e_{3}} & = & -\frac{3}{2}\left( 9z_{q}+ z_{\ell_{1}} \right) - z_{\ell_{2}} ~.
\label{eq:one-e}
\end{eqnarray}  
The remaining lepton-doublet charge, $z_{\ell_3}$, is given by the second equation in (\ref{eq:zl3}).
Two other solutions exist, in which $z_{\ell_1}$ is replaced in all  Eqs.~(\ref{eq:one-n}) by
either $z_{\ell_2}$ or $z_{\ell_3}$, leading to different expressions for $z_{e_i}$ 
which can be computed from Eqs.~(\ref{eq:e2e3}) and (\ref{eq:e1}).
All three solutions yield similar phenomenology. We will discuss in what follows 
only the solution shown in Eqs.~(\ref{eq:one-n}) and (\ref{eq:one-e}).

An important feature of this $N=1$ model is that one of the Dirac mass terms 
is renormalizable. This can be seen by replacing $z_u$ and $z_n$ in 
the first Eq.~(\ref{eq:exp-gen}), which determines the dimensionalities of the operators that 
induce Dirac masses: it turns out that one of the $p_{i1}$ exponents vanishes.
For the solution considered here, $p_{11} = 0$.
As a result, one of the Dirac masses is equal to the electroweak scale times
a Yukawa coupling, and a seesaw mechanism needs to be invoked for explaining 
the smallness of the neutrino masses compared to the other lepton and quark masses.
Furthermore, one of the $q_{ij}$ exponents vanishes ($q_{23} = 0$ for the solution 
considered here), so that one of the left-handed  Majorana masses is 
$v^2/\Lambda$ times a dimensionless coefficient, which suggests $\Lambda\gg v$.
All other exponents shown in Eq.~(\ref{eq:exp-gen}) 
can be expressed in terms of only two parameters,
defined by 
\begin{eqnarray}
&&r\equiv -3\left(3z_q + z_{\ell_1}\right)~, \nonumber\\ [0.2em]
&&\delta \equiv z_{\ell_2}-z_{\ell_1} ~.
\end{eqnarray}
After spontaneous gauge symmetry breaking, the symmetric 4$\times 4$ neutrino mass matrix,
$M_{\nu}$, is given by
\begin{equation}
\left(\begin{array}{ccc|c}\displaystyle
c_\ell^{11} \frac{v^{2}}{\Lambda}\epsilon^{|r|} 
& \displaystyle c_\ell^{12} \frac{v^{2}}{\Lambda} \epsilon^{|r-\delta |}
& \displaystyle c_\ell^{13} \frac{v^{2}}{\Lambda} \epsilon^{ |\delta|}
& \lambda_\nu^{11} v 
\\ [1.2em]
& \displaystyle c_\ell^{22} \frac{v^{2}}{\Lambda} \epsilon^{|r-2\delta |} 
& \displaystyle c_\ell^{23} \frac{v^{2}}{\Lambda} 
& \lambda_\nu^{21} v \, \epsilon^{|\delta |}
\\ [1.2em]
& & \displaystyle c_\ell^{33} \frac{v^{2}}{\Lambda} \epsilon^{|r-2\delta |} 
& \; \lambda_\nu^{31} v \, \epsilon^{|r-\delta |} \\  [1em] \hline &&&\\  [-1em]
&&&  c_n \Lambda \epsilon^{|r|}  \end{array}\right)  ~,
\label{mnu_1}
\end{equation}
where again it is understood that elements proportional to $\epsilon$ raised to a 
non-integer power vanish.
Small enough neutrino masses may be obtained from a seesaw mechanism provided
the $(4,4)$ element is large, so  
we take $r=0$ ($r=\pm 1$ may also work if $\epsilon$ is large enough).
This being the case, the right-handed neutrino is very heavy (as in the  high-energy seesaw) and 
has tiny mixing with the active neutrinos. 
For $\delta=0$ all $U(1)_\nu$ charges are proportional to the corresponding hypercharges,
and the neutrino masses are not trimmed by the new gauge symmetry, so
we concentrate on the less trivial case where  $\delta\neq 0$.

In order to accommodate the large neutrino mixing, one can take advantage of the form of the 
the $M_{\nu}$ matrix in the $\epsilon=0$ limit [see Eq.~(\ref{mnu_1})]. 
That has a $L_{\mu}-L_{\tau}$  lepton flavor 
symmetry which yields three massive neutrino states, two of which are degenerate fifty-fifty admixtures 
of $\nu_{\mu}$ and $\nu_{\tau}$. If $|\delta|$ is a small integer or half integer (like $1/2$ or $1$) 
and $\epsilon$ is large ({\it i.e.} $\epsilon\sim 0.1$), one can hope that such $L_{\mu}-L_{\tau}$ 
breaking effects are sufficient to explain the other lepton mixing features. Unfortunately, reasonable 
choices for $\epsilon$ and $\delta$  yield a too small solar mixing angle. 
We find, however, that this solution can be rescued by the fact that the mixing among 
electrically-charged leptons
can contribute significantly to solar neutrino mixing. 

The charged-lepton mass matrix is given by
\begin{equation}
M_{e}= v \left(\begin{array}{ccc}
\lambda_e^{11} & \lambda_e^{12} \epsilon^{|\delta |} & \lambda_e^{13} \epsilon^{|-r+\delta |}
\\
&&\\
 \lambda_e^{21} \epsilon^{|\delta |} & \lambda_e^{22} & \lambda_e^{23}\epsilon^{|r-2\delta |}
\\
&&\\
  \lambda_e^{31}\epsilon^{|-r+\delta |} &  ~ \lambda_e^{32} \epsilon^{|r-2\delta |} & \lambda_e^{33}
\end{array}\right) ~~.
\label{n=1,cc}
\end{equation} 
It is easy to find values for
the coefficients in the $M_e$ and $M_\nu$  matrices that lead to 
experimentally allowed values for neutrino mixing angles. This is facilitated by the fact that
some of the the dimensionless couplings $\lambda_e$ must be 
very small in order to reproduce the electron mass, so that large mixing angles can arise
after diagonalizing Eq.~(\ref{n=1,cc}). 

Overall, the  $N=1$ model 
is indistinguishable from any scenario 
where neutrino masses are associated with physics at a very high energy scale. Furthermore, given that 
the $U(1)_{\nu}$ breaking scale is expected  to be very high, the $Z^\prime$ boson is expected to be
out of the reach of collider experiments. A large $\langle\phi\rangle$ also implies that the rates for 
charged-lepton flavor violating phenomena are also out of the reach of next-generation experiments,
in spite of the fact that they are mediated at tree level by $Z^\prime$ exchange.  


\section{Cosmological Constraints}
\label{sec:cosmo}

In all scenarios discussed in Sec.~\ref{sec:model}, there are new light fermionic degrees of freedom.
Their presence has non-trivial consequences for cosmology. 
Since right-handed neutrinos 
are charged under $U(1)_{\nu}$, we expect them to reach thermal equilibrium in early times, and only
to fall out of thermal equilibrium at around the same time that active neutrinos fall out of 
thermal equilibrium, assuming that the $U(1)_{\nu}$ symmetry breaking scale is close to the 
electroweak scale. This being the case, the existence of very light ($m\ll 1$~eV) quasi-sterile neutrinos 
is constrained by Big Bang Nucleosynthesis (BBN) and other cosmological data, including precision measurements of the cosmic microwave background and the distribution of mass and energy at very large scales~\cite{Barger:2003zh}. 
The existence of light sterile neutrinos  ($1~{\rm eV}\lesssim m\lesssim  1$~GeV) is also constrained both by BBN and constraints on the amount of matter (hot, warm, or cold) in the Universe. 
We emphasize that even in the case of Dirac neutrinos, the equivalent number of neutrinos $N_{\nu}$ is
larger than three for a low enough $U(1)_{\nu}$ breaking scale. 

Recent analyses of cosmological data (mostly the cosmic microwave background and large scale structure) point to a number of relativistic degrees of freedom equivalent to somewhere between 3 and 6 neutrinos ($3\lesssim N_{\nu}\lesssim 6$) \cite{Cirelli:2006kt,Hannestad:2006mi}. A recent analysis of BBN, on the other hand, constrains $2\lesssim N_{\nu}\lesssim 5$ \cite{Hannestad:2005jj}, while other estimates are much more stringent, $1.7\lesssim N_{\nu}\lesssim 3$  \cite{Barger:2003zg}. The key difference between \cite{Hannestad:2005jj} and  \cite{Barger:2003zg} comes from the assumption regarding the observational value of the primordial $^4$He abundance, including its uncertainty. More specifically, \cite{Hannestad:2005jj} assumes that the most realistic estimate of the $^4$He abundance is the one presented in \cite{Olive:2004kq}, which quotes an uncertainty roughly three times larger than the more widely used result \cite{Steigman:2005uz}.  We conservatively interpret the current data to indicate that one or perhaps two extra equivalent neutrinos are marginally allowed, while three are likely ruled out. 

Assuming that sterile neutrinos are kept in thermal equilibrium by $Z'$ exchange, we can readily estimate that they ought to decouple at $T=T_d$, given roughly  by
\begin{equation}
\frac{T_d^2}{M_{\rm Pl}}\sim \frac{T_d^5}{4\pi\langle\phi\rangle^4}\ ,
\end{equation}
if the $Z'$ mass is much larger than $T_d$. Hence, sterile neutrinos contribute to the expansion rate at the time of BBN like $(10.75/g_*(T_d))^{4/3}$ equivalent neutrinos. For $T_d$ values above the QCD phase-transition (around 180~MeV), $g_*\gtrsim 60$, so that each sterile state contributes like 0.1 neutrinos (or less) and is easily allowed by the data. $T_d\gtrsim180$~MeV translates into $\langle\phi\rangle\gtrsim5$~TeV. 

For smaller $U(1)_{\nu}$ breaking scales there are several ways out. 
Modest modifications to the concordance cosmological model allow for more relativistic degrees of freedom at the time of BBN, including allowing for a large lepton asymmetry among the active leptons. The authors of \cite{Barger:2003rt}, for example, find that up to four massless neutrinos can be added to the primordial universe as long as the electron neutrino chemical potential is $\xi_e\sim 0.2$. 

Heavier sterile neutrinos ($m\gtrsim1$~eV) must also satisfy constraints on the amount of matter  (hot or cold) in the universe. If the sterile neutrinos decouple while relativistic ($m_{\nu_s}\lesssim 100$~MeV for $\langle\phi\rangle\gtrsim 5$~TeV), their contribution to the critical density is estimated to be 
\begin{equation}
\Omega_{\nu_s}\sim 0.2\left(\frac{m_{\nu_s}}{100~\rm eV}\right).
\end{equation}
Hence, mostly right-handed states with masses above 100~eV would 
overclose the universe. A 100~eV sterile neutrino would behave as hot dark matter, whose contribution to the energy budget of the universe is currently constrained to be much less than the estimate above. On the other hand, we estimate that much heavier sterile states ($m\gtrsim 100$~MeV) will decouple while non-relativistic, and serve as good dark matter candidates if their masses are above tens of GeV, and otherwise overclose the universe. 

In summary, very light quasi-sterile neutrinos ($m\lesssim 10$~eV) are in agreement with early universe data if $\langle\phi\rangle\gtrsim 5$~TeV, even if we do not appeal to non-standard cosmology. Smaller $\langle\phi\rangle$ are easily allowed if one adds new ingredients to the early universe, like a large lepton asymmetry.  Heavier mostly sterile states ($10~{\rm eV}\lesssim m\lesssim 10$~GeV) either populate the universe with too much hot dark matter or too much matter ($\Omega_{\nu_s}\gg 1$). These  constraints can be circumvented in a variety of ways, including adding new sterile neutrino interactions that will keep the heavy states in thermal equilibrium until lower temperatures, or postulating a low reheating temperature ($T_{\rm reheat}\lesssim 100$~MeV, easily allowed by current data \cite{Hannestad:2004px}, should suffice).

\section{Collider probes of neutrino mass generation}
\label{sec:collider}

If the $U(1)_\nu$ gauge symmetry which controls the higher-dimensional operators 
responsible for generating neutrino masses is spontaneously broken at or below 
the TeV scale, then the associated $Z^\prime$ gauge boson is likely to produce 
observable effects at high-energy colliders. Here we discuss the case where 
the gauge coupling is not much smaller than unity, so that the $Z^\prime$
boson may be produced copiously at the LHC~\cite{Rizzo:2006nw}. Furthermore, we assume that 
the $Z^\prime$  mass is below 1 TeV so that it can show up 
as a resonance at the ILC. 

For nonzero values of the $U(1)_\nu$ charge of the Higgs doublet, there is 
tree-level mixing between the $Z$ and $Z^\prime$ bosons, which is tightly 
constrained by the LEPI data (see Fig. 1 of Ref.~\cite{Carena:2004xs}). 
We will thus consider only the $z_H=0$ case. In Secs. III and IV we 
have studied several $U(1)_\nu$ charge assignments. 
The most general one consistent with neutrino mass generation at the TeV-scale  
is that of the 2+1 Leptocratic Model (note that the Orwellian Leptocratic
Model is a particular case with $a=0$). Imposing the additional condition of 
$z_H=0$ we find 
\be
c= 6 z_q ~,
\ee
so that all $U(1)_\nu$ charges are given in terms of only two rational
parameters, $a$ and $b$. It is convenient to normalize the gauge coupling such that 
the quarks have $U(1)_\nu$ charge $+1/3$. The other $U(1)_\nu$ charges are listed in
Table II. 

\begin{table}[t]
\renewcommand{\arraystretch}{1.5}
\begin{tabular}{|c|c|}\hline 
\ field \ & \ $U(1)_\nu$ charge \ \\  [0.03em] \hline\hline & \\ [-0.7em]
$q_L$,  $u_R$, $d_R$ & \ $\displaystyle \frac{1}{3}$ \ \\ [0.5em]
$\ell_L^1$, $e_R^1$ & \ $-1 - 2 a z_\phi $ \ \\ [0.3em]
$\; \ell_L^2$, $\ell_L^3$, $e_R^2$, $e_R^3 \;$ & \ $- 1 + a z_\phi$ \ \\ [0.3em]
$n_R^1$ & \ $-1-2 b z_\phi$ \ \\  [0.3em]
\ $n_R^2, n_R^3$ \ & \ $-1 + b z_\phi $ \ \\  [0.3em] \hline
$H$ & $0$ \ \\ [0.2em]
$\phi$ & $\; \displaystyle z_\phi = - \frac{3(a+b)}{a^2+ab+b^2} \; $ \\ [0.6em] \hline
\end{tabular}
\medskip \\
\vspace{0.2in}
\caption{\small The two-parameter family of $U(1)_\nu$ 
charges 
in  the 2+1 Leptocratic Model ($a^\prime = b^\prime = 0$) 
with the additional constraint of 
$z_H=0$. The Orwellian Leptocratic Model with $z_H=0$ is recovered for $a=0$.}
\end{table}

In the event of a $Z^\prime$ discovery in  dilepton channels at the 
LHC or the Tevatron, it would be  straightforward to measure the ratio of branching fractions 
into $e^+e^-$ and $\mu^+\mu^-$. Unlike the majority of models studied in the
literature, the 2+1 Leptocratic Model with $a\neq 0$ predicts a value for this ratio
different than unity:
\be
\frac{B\left( Z^\prime \rightarrow e^+e^-\right)}
{B\left( Z^\prime \rightarrow \mu^+\mu^-\right)}
= \left( \frac{1 + 2 a z_\phi}{1 - a z_\phi} \right)^2  ~,
\ee
where the charge of the scalar field $\phi$ is given by
\be
z_\phi = - \frac{3(a+b)}{a^2+ab+b^2} ~.
\ee
Measuring this ratio would allow the extraction of the $az_\phi$ combination of 
parameters. If furthermore a resonance of the same invariant mass is discovered
in the $t\overline{t}$ channel at the LHC or the Tevatron, then it will be 
straightforward to test the 2+1 Leptocratic Model which predicts
\be
\frac{B\left( Z^\prime \rightarrow e^+e^-\right)}
{B\left( Z^\prime \rightarrow t\overline{t}\right)}
= 3\left( 1+2 a z_\phi \right)^2  ~.
\ee

Let us assume now that a resonance will be discovered in dilepton channels,
and that the ratio of branching fractions into $e^+e^-$ and $\mu^+\mu^-$
turns out to be equal to one within experimental 
errors. The Orwellian Leptocratic Model ($a=0$) will be favored over 
the more general 2+1 Leptocratic Model. The question is how to establish that 
the resonance is indeed associated with our $U(1)_\nu$ and not some
other extension of the standard model. 
Let us assume that the ATLAS and CMS experiments at the LHC will be able to determine
precisely several properties of the $Z^\prime$ boson by measuring total rates, 
angular distributions, and other observables in the $e^+e^-$, $\mu^+\mu^-$, 
$t\overline{t}$ and perhaps a couple of other channels,  such that all the results 
are consistent with the Orwellian Leptocratic Model. Will that be enough 
evidence that the neutrino masses are generated at the TeV scale rather than some 
very high seesaw scale? The answer is no, because in the Orwellian Leptocratic Model
all standard model fermions have charges given by their $B-L$ number if $z_{H} = 0$.
It turns out that extending the electroweak gauge group to 
$SU(2)_L\times U(1)_Y\times U(1)_{B-L}$ is a natural possibility for TeV scale physics
which does not lead to an explanation for the smallness of the neutrino masses.
The only way to distinguish experimentally between the $Z^\prime$ from the  
Orwellian Leptocratic Model and the $Z_{B-L}$ boson associated with 
the $U(1)_{B-L}$ gauge symmetry is by measuring the branching fraction for
invisible decays. 

In the Orwellian Leptocratic Model we find
\be
B\left( Z^\prime \rightarrow \, {\rm invisible} \, \right) = \frac{6}{7} ~,
\ee
where we ignored the top mass, and assumed that the decay into the 
CP-even component of the $\phi$ scalar is kinematically forbidden.
This large invisible branching fraction is a consequence of the large 
charges of the right-handed neutrinos: $z_{n_{1}}=5$ and $z_{n_{2}}=z_{n_{3}} = -4$ in the normalization
where the quarks have charge 1/3 and the leptons have charge $-1$.
The branching fraction for invisible decays of the $Z_{B-L}$ boson
is significantly smaller, given by $6/16$.

A measurement of the invisible decay of a $Z^\prime$ boson at the LHC 
would be extremely hard. For triggering purposes, the $Z^\prime$ would have to be produced in 
association with some other particles,  which would render  the signal rates small.  
At the same time, the backgrounds are likely
to be large. The best hope for measuring the invisible decay of a $Z^\prime$ boson
is provided by the ILC, where the total production cross section is
well known \cite{Freitas:2004hq}, and the backgrounds will be under control.

The scalar sector responsible for $U(1)_\nu$ breaking may also be accessible at 
colliders. Assuming that a single $\phi$ scalar is charged under $U(1)_\nu$,
its CP-even degree of freedom may be produced in association with the 
$Z^\prime$ boson. Based on the structure of the operators responsible for neutrino masses,
its main decay mode would be into neutrinos. A more interesting channel, albeit 
with a phase-space suppressed branching fraction, is into a charged lepton, a longitudinal 
$W$ boson and a sterile neutrino. We point out, though, that besides the operators responsible
for neutrino masses, other higher-dimensional operators may lead to large branching fractions
of the $\phi$ scalar into quarks and charged leptons, and possibly into standard model Higgs bosons. 
For example, the gauge-invariant 
dimension-six operator 
\be
\frac{1}{\Lambda^2}\phi^\dagger \phi \, (\overline{t}_L, \overline{b}_L) \tilde{H} t_R 
\ee
may lead to a dominant $\phi$ decay into top quarks. Hence, the phenomenology 
of the $U(1)_\nu$-breaking sector is more model-dependent than that of the 
$Z^\prime$ boson. An interesting possibility is to check whether there are more 
$\phi$ scalars coupled to the $Z^\prime$, which would further test the 
operators responsible for neutrino masses.

\setcounter{equation}{0}
\setcounter{footnote}{0}
\section{Summary and Conclusions}
\label{sec:end}
The most popular explanation for tiny neutrino masses is to postulate that lepton number is a symmetry of the standard model that is broken at
an energy scale $\Lambda$, close to the grand unification scale.
Besides suppressing the neutrino masses by $v/\Lambda$,
the high  energy versions of the seesaw mechanism provide all necessary ingredients
to explain the matter-antimatter asymmetry of the universe \cite{leptogenesis}.
On the more sobering side, a very high energy origin for neutrino masses cannot be verified
experimentally; one can at most envision accumulating indirect evidence for the
physics behind neutrinos masses \cite{Buckley:2006nv}.

Here, we have pursued a different approach. We investigated the generation of neutrino
masses in a non-anomalous $U(1)_{\nu}$-extended standard model. Its particle content
includes $N$ right-handed neutrinos, which are neutral under $SU(3)_c\times SU(2)_L\times U(1)_Y$, but have non-trivial charges under the $U(1)_{\nu}$ symmetry.
Right-handed neutrinos allow for non-trivial, non-anomalous
extensions of the gauge sector, while the $U(1)_{\nu}$ provides a natural mechanism for
generating small neutrino masses that does not necessarily
rely on physics at energy scales significantly above
the electroweak scale. Generically,
$U(1)_{\nu}$ gauge invariance forbids the usual
neutrino mass terms, and  these are  generated only through operators of high mass dimension
which include scalar fields associated to the $U(1)_{\nu}$ breaking scale $\langle\phi\rangle$.
Hence,
neutrinos are light because their masses are suppressed by several powers of $\langle\phi\rangle/\Lambda$. This suppression factor is small 
even for an ultraviolet cutoff $\Lambda$ close to the TeV scale and $\langle\phi\rangle$ of order the 
electroweak scale.

The new anomaly cancellation conditions
are highly non-trivial, especially because all fermion
charges are expected to be commensurate. Nevertheless, assuming that all quark Yukawa couplings
and all diagonal charged-lepton Yukawa couplings to the standard model Higgs doublet $H$ are gauge invariant,
we have found the most general solution to the anomaly cancellation conditions when
$N=1$ or 3. Only in the  $N=3$ case were we able to find scenarios consistent with
light neutrino masses and
$\Lambda$ at the TeV scale, unless the Lagrangian contains very small dimensionless coefficients
(smaller than the standard model electron Yukawa coupling).
For $N=3$, the charges of all quarks and leptons
(including right-handed neutrinos) are determined in terms of four rational parameters,
assumig one of the fermion charges is fixed by an appropriate normalization of the gauge
coupling. This four-parameter charge assignment, referred to as the Leptocratic Model, is
shown in Eqs.~(\ref{leptocratic-1})-(\ref{leptocratic-5}) and (\ref{cab}).

Even when only one scalar field other than $H$ has a  $U(1)_{\nu}$-breaking VEV, we have
found regions in the Leptocratic Model  parameter space that fit the neutrino oscillation data.
These are discussed in detail in Sec.~\ref{sec:model}.
Depending on the choice of parameters, the neutrinos can be  either
Dirac or Majorana fermions.
In scenarios with Majorana neutrinos, we predict the existence of
``quasi-sterile'' neutrinos that mix slightly with the active neutrinos and couple to the new
$Z^{\prime}$ gauge boson. These quasi-sterile neutrinos may have interesting
phenomenological consequences 
for cosmology and oscillation physics. In the case of Dirac neutrinos, we also predict
potentially observable consequences of the new degrees of freedom.

Since the $U(1)_{\nu}$ symmetry is spontaneously broken around the weak scale, 
the $Z^{\prime}$ gauge boson and the particles from the $U(1)_{\nu}$ breaking sector
will manifest themselves in a variety of interesting ways.  
$Z^{\prime}$ exchange can mediate neutral-fermion flavor
violating processes such as $\nu_e X\to \nu_{\tau}X$, which may be observable in next-generation
neutrino oscillation experiments. The new heavy states can be discovered
in current and upcoming collider experiments. If this turns out to be the case, the
Tevatron, LHC and ILC may provide our first direct glimpse of the physics unveiled by neutrino oscillation
experiments.

\section*{Acknowledgments}

The work of AdG is sponsored in part by the US Department of Energy Contract DE-FG02-91ER40684.
Fermilab is supported by the US Department of Energy Contract DE-AC02-76CHO3000.

\appendix*
\section{Active-Sterile mixing in Orwellian Leptocratic Model}
\label{app:mixing}
\renewcommand{\theequation}{A.\arabic{equation}}

In the Orwellian Leptocratic Model discussed in Sec.~\ref{sec:Orwellian}, if $b$ is an integer multiple of $3/2$ there exist light sterile neutrinos which mix with the active ones. In this appendix, we derive the mixing among these states. 

If $b$ is a multiple of $3$ (this corresponds to the second Orwellian case), all elements in the Dirac and Majorana mass matrices are allowed. 
The right-handed Majorana mass matrix gives a light eigenstate with 
mass $\epsilon^{8|b|/3} \Lambda$ and two heavy eigenstates with mass $\epsilon^{|b|/3} \Lambda$. The mixing matrix that diagonalizes the right-handed neutrino mass matrix is ``bi-large.''  Hence, in the basis where the right-handed neutrino  mass matrix is diagonal, the  Dirac neutrino mass matrix is anarchical, and all its entries are of order $v\epsilon^{|b|}$. In this basis, the $6 \times 6$ 
mass matrix, $M_\nu$, has the following form:
\begin{equation} 
\left(\begin{array}{cc}
  \displaystyle
\frac{v^{2}}{\Lambda} \, \epsilon^{2|b|/3} (c_{\ell})_{\scriptscriptstyle  3\times 3}
&  v \epsilon^{|b|} \, (\lambda^{\prime}_{\nu})_{\scriptscriptstyle 3 \times 3} 
\\ [1em]
v \epsilon^{|b|} \, (\lambda^{\prime}_{\nu})_{\scriptscriptstyle 3 \times 3}^\top 
&  \;\;\Lambda \epsilon^{|b|/3} \, {\rm diag}\left( c^{\prime11}_{n} \epsilon^{7|b|/3},\;
c^{\prime22}_{n},\,  c^{\prime33}_{n} \right)
\\ \end{array}\right) \; .
\label{big-matrix}
\end{equation}
Here, $\lambda^{\prime}_{\nu}$ is a $3\times 3$ matrix whose entries are linear combinations of $\lambda^{ij}_{\nu}$ 
and $c_{n}^{ij}$, 
$c^{\prime}_{n}$ are linear combinations of $c_n^{ij}$, and diag denotes a diagonal matrix.

Two of the three masses of the mostly right-handed neutrino states are much larger than the Dirac masses, 
and can be described by a $5 \times 5$ effective mass matrix where we keep the active neutrinos and 
only the ``second'' and the ``third'' sterile neutrinos
({\it i.e.}, we discard the 4th row and 4th column of the above $M_\nu$ matrix).
Two of the mostly active neutrinos receive a Type-I seesaw contribution to their masses
 \begin{equation}
 m_{\nu}^{\rm seesaw} \sim 
\frac{\lambda_{\nu}^{2}}{c_{n}} \frac{v^{2}}{\Lambda} \epsilon^{5|b|/3} \; .
 \end{equation}
 Here and in the remainder of this Appendix, $\lambda_{\nu}, \; c_{n}$ and $c_{\ell}$ stand for typical elements of the respective $3\times 3$ matrices. 
These contributions, however, are negligible, as they are much smaller than the left-handed Majorana masses.  
As a result, the mostly active neutrino masses come predominantly from the left-handed Majorana mass matrix, which gives three equal-magnitude masses for the active neutrinos,
 \begin{equation}
 m_{\nu_{1,2,3}} \sim c_{\ell} \epsilon^{2|b|/3} \frac{v^{2}}{\Lambda}
 \; .
 \end{equation}
 The observed large lepton mixing comes from the anarchical structure of the left-handed Majorana mass matrix. 
 
 The two heavy sterile neutrino  masses are not affected by the seesaw,
 \begin{equation}
 m_{\nu_{5,6}} \sim c_{n} \Lambda \epsilon^{|b|/3}
 \end{equation}
 and the mixing between the two heavy sterile neutrinos and the active ones is,
 \begin{equation}
 \Theta_{\rm active-heavy} 
 \sim \frac{\lambda_{\nu}}{c_{n}} \epsilon^{2|b|/3} \frac{v}{\Lambda}
\; .
 \end{equation}  
 
The active-light sterile neutrino mass matrix is given by a  $4 \times 4$ 
effective mass matrix where we keep the active neutrinos and only the ``first'' sterile neutrino
[{\it i.e.}, we discard the 5th and 6th rows and columns of the $M_\nu$ matrix
given in Eq.~(\ref{big-matrix})].
Given that 
\be
c_{n} \Lambda \epsilon^{8|b|/3} \ll c_{\ell} \frac{v^{2}}{\Lambda} \epsilon^{2|b|/3}, 
\; \lambda_{\nu} v \epsilon^{|b|/3} ~,
\ee
further diagonalizing this matrix, we obtain the lightest sterile neutrino mass from an ``inverted seesaw'' contribution,
\begin{equation}
m_{\nu_{4}} 
\sim \frac{\lambda_{\nu}^{2}}{c_{\ell}} \Lambda \epsilon^{4|b|/3} \; ,
\end{equation}
while the mixing between the active and the lightest sterile neutrinos is given by,
\begin{equation}
\Theta_{\rm active-light} 
\sim \frac{\lambda_{\nu}}{c_{\ell}} \epsilon^{|b|/3} \frac{\Lambda}{v} \; .
\end{equation} 

If $b$ is an odd multiple of $3/2$, which corresponds to the third Orwellian case, the full $6\times 6$ seesaw matrix is of the following block-diagonal form 
\begin{equation}
\left(\begin{array}{ccc}
\displaystyle{\frac{v^{2}}{\Lambda}} \epsilon^{2|b|/3} (c_{\ell})_{\scriptscriptstyle 3 \times 3}  & 
 v \epsilon^{2|b|} (\lambda_{\nu})_{\scriptscriptstyle 3\times 1} & 0_{\scriptscriptstyle 3\times 2}
\\ [1em]
& c_{n}^{11} \Lambda \epsilon^{10|b|/3} \quad & 0_{\scriptscriptstyle 1 \times 2}
\\
&&\\
& &  \!\!\!\!\!\!\!\Lambda \epsilon^{8|b|/3}   {\rm diag}\left( c^{\prime22}_{n},\,  c^{\prime33}_{n} \right)
\end{array}\right) ~,
\end{equation}
as some of the matrix elements are forbidden by gauge invariance (because the matrix is symmetric, we have omitted 
the lower off-diagonal elements).  
The diagonalization of the right-handed neutrino mass matrix reveals the following eigenvalues: two heavier ones with masses of order $\sim \epsilon^{8|b|/3} \Lambda$, and the lightest one with mass $\sim \epsilon^{10|b|/3} \Lambda$. As these masses are much lighter than the Dirac masses, there is no usual seesaw contribution to the active neutrino masses. Furthermore, due to the block-diagonal form of the full $6\times 6$ mass matrix, the heavier two sterile neutrinos do not mix with the active ones. The three active neutrinos acquire comparable masses from the left-handed Majorana mass matrix,
\begin{equation}
m_{\nu_{1,2,3}} \sim c_{\ell}\frac{v^{2}}{\Lambda} \epsilon^{2|b|/3} \; ,
\end{equation}
since the left-handed Majorana masses are much larger compared to the Dirac masses. The lightest sterile neutrino acquires its mass through the following inverted seesaw contribution,
\begin{eqnarray}
m_{\nu_{4}} 
\sim \frac{\lambda_{\nu}^{2}}{c_{\ell}} \epsilon^{10|b|/3} \Lambda \; .
\\
\nonumber
\end{eqnarray}
The mixing between the active neutrinos and the lightest sterile one is,
\begin{equation}
\Theta_{\rm active-light} 
\sim \frac{\lambda_{\nu}}{c_{\ell}} \epsilon^{4|b|/3} \frac{\Lambda}{v} \; .
\end{equation} 
The active-sterile mixing in this case is significantly smaller compared to the previous case.


 
 \end{document}